\documentclass[aps,pre,twocolumn,showpacs,notitlepage,superscriptaddress]{revtex4-1}

\usepackage{amsmath}
\usepackage{amssymb}
\usepackage{mathtools}
\usepackage{titlesec}
\usepackage[all]{xy}
\input xy
\xyoption{matrix}
\xyoption{2cell}
\xyoption{arrow}
\xyoption{curve}
\xyoption{all}

\usepackage{hyperref}
\usepackage{amsmath}
\usepackage{graphicx}
\usepackage[latin1]{inputenc}
\usepackage{ulem}
\usepackage{epstopdf}
\usepackage{subfigure}
\usepackage{xcolor}

\titlespacing\subsection{0pt}{8pt}{8pt}

\newcommand{\abs}[1]{\left| #1 \right|}
\newcommand{\ie}{\textit{i.e.}}
\begin{document}

%%%%%%%%%%%%%%%%%%%%%%%%%%%%%%%%%%%%%%%%%%%%%%%%%%%%%%%%%%%%%%%%%%%%%%%%%%%%%
\title{Flatband Engineering of Mobility Edges}
\author{Carlo Danieli}
\affiliation{New Zealand Institute for Advanced Study, Centre for Theoretical Chemistry \& Physics, Massey University, Auckland, New Zealand}
\author{Joshua D. Bodyfelt}
\affiliation{New Zealand Institute for Advanced Study, Centre for Theoretical Chemistry \& Physics, Massey University, Auckland, New Zealand}
\author{Sergej Flach}
\affiliation{New Zealand Institute for Advanced Study, Centre for Theoretical Chemistry \& Physics, Massey University, Auckland, New Zealand}
\affiliation{Center for Theoretical Physics of Complex Systems, Institute for Basic Science, Daejeon, Korea}
\begin{abstract}
Properly modulated flatband lattices have a divergent density of states at the flatband energy. Quasiperiodic modulations are known to host a metal insulator transition already
in one space dimension.  Their embedding into flatband geometries consequently allows for a precise engineering and fine tuning of mobility edges. We obtain analytic expressions for
singular mobility edges for two flatband lattice examples. In particular, 
we engineer cases with arbitrarily small energy separations of mobility edge, 
zeroes, and divergencies.  
\end{abstract}
\maketitle
%%%%%%%%%%%%%%%%%%%%%%%%%%%%%%%%%%%%%%%%%%%%%%%%%%%%%%%%%%%%%%%%%%%%%%%%%%%
%%
\section{Introduction}
The phenomenon of wave localization has been intensively studied since its prediction in 1958 \cite{Anderson58}, where
complete localization was proved in the case of a one-dimensional (1-D) chain defined over a random potential. 
Moreover, it was shown that the 3-D case allows for an energy-dependent transition from localized to delocalized eigenstates. The transition has been since coined
metal-insulator transition (MIT). The critical energy $E_c$ is called a
mobility edge; in general, it depends on and varies upon changes of the control 
parameters of the given model  \cite{Bulka85}.
Interestingly, an MIT can also be realized in one-dimensional settings 
with sufficiently correlated disorder potentials \cite{Izrailev99}.

In 1980, Aubry and Andr\'e proved the existence of the MIT for a 1-D chain 
defined over a specific quasiperiodic potential \cite{Aubry80}. This MIT occurs 
at a critical value ($\lambda_c=2$) of the onsite potential's strength 
$\lambda$, and separates the metallic phase $\lambda\in]0,2[$ from an insulating 
phase $\lambda\in]2,+\infty[$. This remarkable result was fully understood via 
the principle of duality, in which a particular Fourier transformation relating 
eigenmodes and energy spectra allows for a direct functional equivalency 
between momentum space and its transform counterpart. This equivalency is 
energy-independent: upon 
crossing the critical value $\lambda_c$ all eigenstates turn from localized to 
extended, regardless of their eigenenergy. The appearance of a mobility 
edge is thus avoided.
Analytic results have been discovered in the last decade regarding the 
topological Cantor structure of the spectrum \cite{Avila09} and its Lebesgue 
measure \cite{Jitomirskaya02}. Furthermore, for each different regime 
(insulating, metallic and critical) different spectral decompositions have been 
found \cite{Avila08, Jitomirskaya99, Gordon97}. Model generalizations were 
reported, e.g. quasiperiodic systems constantly maintained at criticality 
\cite{ostlund83, kohmoto83}, bichromatic quasiperiodic lattices displaying 
mobility edges \cite{hiramoto89, boers07}, and completely localized 
quasiperiodic models \cite{grempel82}. Correlated metallic states have also been 
observed in the insulating regime, for the case of two interacting particles 
within a 1-D Aubry-Andr\'e chain \cite{Flach12}. 
In another recent work \cite{ganeshan2014}, a suitably modified quasiperiodic 
potential was shown to produce a mobility edge expressable in an analytic form - 
a property which we will take to new limits using flatband topologies.

Wave propagation on lattices with flatband topologies is characterized by the 
existence of horizontal (flat) bands in their band structure. Known in condensed 
matter, this model class has gained great interest in the scientific community, 
due in part to experimental realizations in optical lattices and paraxially 
approximate light propagations \cite{bloch05, christodoulides03, masumoto12}. 
Recent theoretical discoveries have also considered the presence of a disordered 
potential \cite{derzhko06, derzhko10} and nonlinearity \cite{leykem13}. An 
innovative procedure detangles flatband states from dispersive ones 
\cite{flach14}. This allows one to inspect specific features of the models as 
they relate to the choice of the onsite perturbations, and also suggests 
specific potential correlations. 
In the present work, this detangling technique of local rotations \cite{flach14} 
is applied as an extension of \cite{bodyfelt14}; in particular, regarding the 
preliminary finding of a MIT occurring in a flatband lattice under quasiperiodic 
Aubry-Andr\'e perturbation.

The present paper has the following structure: in Sec.\ref{sec:topos} the 
general features of flatband topologies are introduced that define two 
particular models (cross-stitch and diamond lattices), a quasiperiodic 
Aubry-Andr\'e onsite perturbation is defined, and the coordinate 
transformation that allows rotation into Fano defect lattices \cite{flach14}. In 
Sec.\ref{sec:cross}-\ref{sec:diamond}, our findings for the cross-stitch and 
diamond lattices are respectively presented: for both, two distinct chain 
correlations are discussed. Where applicable, the exact mathematical expression 
obtained for the mobility edge is jointly shown with numerically obtained 
transitions for these particular onsite correlations. 
%%%%%
\section{Flatband Topologies}\label{sec:topos}
%%%%%
Consider the eigenvalue problem of a generalized tight-binding model
\begin{equation}
E\psi_n = \epsilon_n\psi_n - \hat{V}\psi_n - \hat{T}(\psi_{n-1} + \psi_{n+1})\ .
\label{eq:fb-gen}
\end{equation}
For all $n\in\mathbb{Z}$, each component of the vector $\psi_n = (\psi_n^1 , \dots,\psi_n^\ell)^T$ represents a site of a periodic lattice, while the set of sites represented by $\psi_n$ is the $n$-th unit cell. The real matrix $ \hat{V}$ defines the geometry of the unit cell, while the real matrix $ \hat{T}$ describes hopping to neighboring cells. At each of the lattice's $i$-th leg $\{\psi_n^i  \}_{n}$, an onsite perturbation $\{  \epsilon_n^i\}_n$ is defined.  The unit cell perturbation $\epsilon_n$ of Eq.(\ref{eq:fb-gen}) is thus given by the diagonal square matrix 
$ \epsilon_n= \mbox{diag}\left( \epsilon_n^a, \epsilon_n^b, \dots, \epsilon_n^\ell \right)$. 

The model geometry is contained in the matrices $ \hat{T}$ and $ \hat{V}$, which are then used to derive the dispersion spectrum via the Bloch solution $\psi_n =\phi_k e^{ikn}$ on an unperturbed crystal $\epsilon_n=0$. Flatband topologies are models in which this crystalline case exhibits \textit{at least} one band independent of $k$ -- such a band is dispersionless, or ``flat''. Eigenmodes corresponding to this flatband energy are (usually) compact localized states (CLS), {\ie}  modes whose amplitude is nonzero only across a finite number of sites
\cite{flach14}. The flatband topology class, $U$, is then defined as the minimum number of unit cells the CLS occupies \cite{flach14}. 

\begin{figure}[h!]
\centering
\includegraphics[width=0.85\columnwidth]{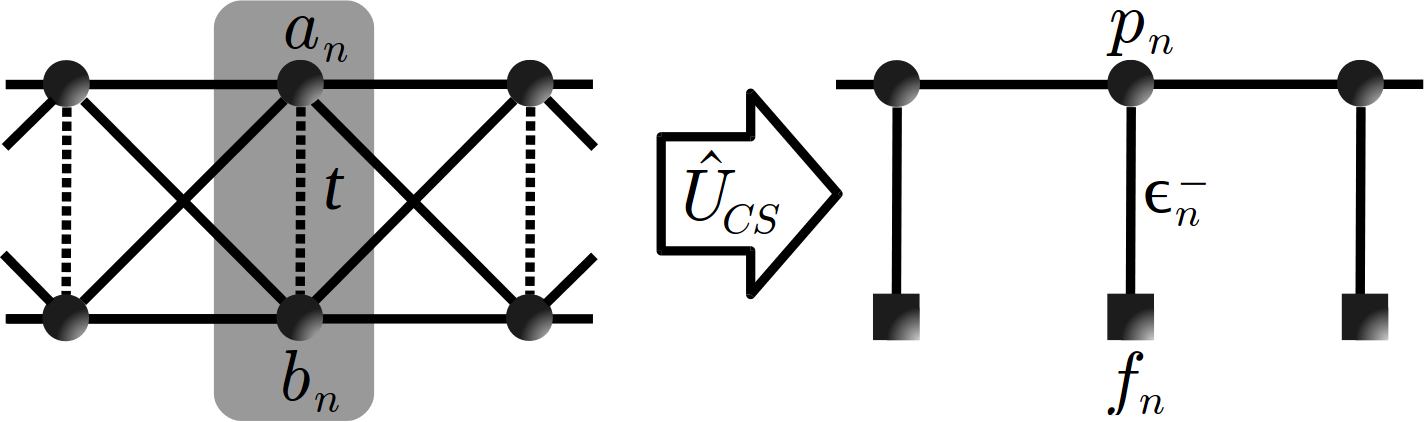}
\caption{Left: The cross-stitch lattice -- the grey shaded region indicates the unit cell. Right: The transformed Fano defect lattice of Eq.(\ref{eq:CS}).}
\label{fig:stitch}
\end{figure}

In this paper we consider two lattice topologies -- the cross-stitch and diamond lattices. The former, shown in the upper left of Fig.\ref{fig:stitch}, is defined for a unit cell $\psi_n=(a_n,b_n)^T$ with a $2\times 2$ peturbation matrix $\epsilon_n$. 
This yields for Eq.(\ref{eq:fb-gen}) the following matrices
\begin{equation}
\hat{V}_{CS} = \left(
\begin{array}{ccc}
0 & t \\
t & 0  
\end{array} \right),\quad
\hat{T}_{CS} = \left(
\begin{array}{ccc}
1 & 1 \\
1 & 1  
\end{array} \right)\;.
\label{eq:CSgeo}
\end{equation}
%%%%%%%%%%%%%%%%%%%%%%%%%%%%%%%%%%%%%%%%%%%%%%%%%%%%%%%%%
%%%%%%%%%%%%%%%%%%%%%%%%%%%%%%%%%%%%%%%%%%%%%%%%%%%%%%%
%%

\begin{figure}[h!]
\centering
\includegraphics[width=0.85\columnwidth]{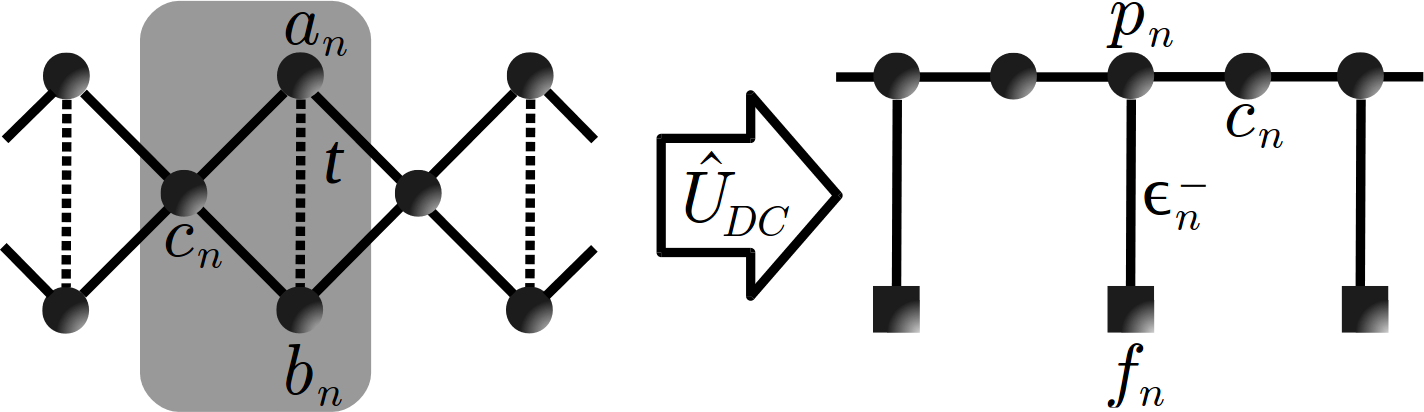}
\caption{Left: The diamond lattice -- the grey shaded region indicates the unit 
cell. Right: The transformed Fano defect lattice of Eq.(\ref{eq:DC}).} 
\label{fig:diamond}
\end{figure}

Likewise for the diamond lattice, as shown in the upper left of Fig.\ref{fig:diamond}, the unit cell is $\psi_n=(a_n,b_n,c_n)^T$ with a $3\times 3$ perturbation matrix $\epsilon_n$. In this case, the matrices in Eq.(\ref{eq:fb-gen}) are
\begin{equation}
\hat{V}_{DC} = \left(
\begin{array}{ccc}
0 & t & 1 \\
t & 0 & 1 \\
1 & 1 & 0 
\end{array} \right),\quad
\hat{T}_{DC} = \left(
\begin{array}{ccc}
0 & 0 & 0 \\
0 & 0 & 0 \\
1 & 1 & 0 
\end{array} \right)\;.
\label{eq:DCgeo}
\end{equation}
%%%%%%%%%%%%%%%%%%%%%%%%%%%%%%%%%%%%%%%%%%%%%%%%%%%%%%%

%%%%%%%%%%%%%%%%%%%%%%%%%%%%%%%%%%%%%%%%%%%%%%%%%%%%%%%

In the unperturbed crystal $\epsilon_n=0$, the dispersive bands are 
\begin{equation*}
E(k) = 
\begin{cases}
- t - 4 \cos k, & \mbox{Cross-Stitch},  \\ 
-\dfrac{1}{2}\left( t \pm \sqrt{t^2 + 16 \cos k + 16}\right), &\mbox{Diamond}.
\end{cases}
\end{equation*}
Additionally, both models contain a flat band at $E(k) = t$.

Associated with the flatband energy, a resulting CLS can be constructed: 
$\psi_n=(1,-1)^T \delta_{n,n_0} / \sqrt{2}$ (cross-stitch) and 
$\psi_n=(1,-1,0)^T \delta_{n,n_0} / \sqrt{2}$ (diamond). Note that both CLS 
are 
contained within a single unit cell. Therefore, according to the definition 
previously stated, both lattices are flatband models class $U=1$. Coordinate 
transformations local to the unit cells rotate these lattices into a Fano defect 
form \cite{flach14}. For the cross-stitch, the rotation is defined by the real 
matrix ${\hat U}_{CS}$ 
\begin{equation}
\left(
\begin{array}{ccc}
p_n \\
f_n  
\end{array} \right) = \hat{U}_{CS}\; \psi_n ,\quad  \hat{U}_{CS} = 
\frac{1}{\sqrt{2}}\left(
\begin{array}{ccc}
1 & 1 \\
1 & -1  
\end{array} \right).
\label{eq:fano-CS}
%\end{split}
\end{equation}
Similarly for the diamond lattice the transformation is defined by the real matrix $U_{DC}$ 
\begin{equation}
\left(
\begin{array}{ccc}
p_n \\
f_n \\
c_n 
\end{array} \right) = \hat{U}_{DC} \; \psi_n , \quad
\hat{U}_{DC} = \frac{1}{\sqrt{2}}\left(
\begin{array}{ccc}
1 & 1 & 0\\
1 & -1 & 0 \\
0 & 0 & \sqrt{2}
\end{array} \right).
\label{eq:fano-DC}
\end{equation}
Lastly, this local coordinate transformation must also rotate the onsite perturbation. For both lattices, this gives
\begin{equation}
\epsilon_n^{\pm}=(\epsilon_n^a \pm \epsilon_n^b)/2.
\label{eq:eps+-}
\end{equation}

The effect of quasiperiodic Aubry-Andr\'e perturbations on these two topologies is the focus of the present work. For both lattices, the onsite perturbations $\{ \epsilon_n^i \}$ are defined as \textit{independent} Aubry-Andr\'e potentials 
\begin{equation}
\epsilon_n^i= \lambda_i \cos\left[ 2\pi \left(\alpha n + \theta_i \right) \right]\ ,
\label{eq:AApot}
\end{equation}
for the $i=a,b$ (cross-stitch) and $i=a,b,c$ (diamond) legs. The parameters $\lambda_i$ are positive real values controlling the perturbative strength, $\theta_i$ is the phase-shift, and $\alpha$ is an irrational number (here set to the golden ratio) called the \textit{incommensurate parameter}. Without loss of generality, the $a$-leg phase can be zeroed ($\theta_a=0$). 
We also set the leg potential strengths equal to each other $\lambda_i = \lambda$.

From Eq.(\ref{eq:eps+-}), notable correlations between the $a$-leg and $b$-leg perturbations appear and will be object of our studies for both models; namely 
\begin{equation}
\begin{aligned}
\mbox{Symmetric:} \quad \epsilon_n^- = 0 & \Leftrightarrow &  \epsilon_n^a = \epsilon_n^b\;, \\
\mbox{Antisymmetric:} \quad \epsilon_n^+ = 0 & \Leftrightarrow &  \epsilon_n^a = -\epsilon_n^b\;.
\end{aligned}
\label{eq:symAsym}
\end{equation}
Since the $a$-leg phase has been zeroed, from Eq.(\ref{eq:AApot}) these two 
correlations are obtained solely from the $b$-leg phase, e.g. $\theta_b=0.5$ 
($\theta_b=0$) for the antisymmetric (symmetric) case. We start the analysis of 
these models with the cross-stitch in 
Sec.\ref{sec:cross}, and then with the diamond lattice in 
Sec.\ref{sec:diamond}.
\section{Cross-Stitch Lattice}\label{sec:cross}
By Eqs.(\ref{eq:fano-CS},\ref{eq:eps+-}), the cross-stitch lattice 
transforms into
\begin{equation}
\begin{aligned}
(E+t)\, p_n &= \epsilon_n^+\, p_n + \epsilon_n^-\, f_n - 2\left( p_{n-1} + p_{n+1} \right)\;, \\
(E-t)\, f_n &= \epsilon_n^+\, f_n + \epsilon_n^-\, p_n\;.
\end{aligned}
\label{eq:CS}
\end{equation}
This results in a Fano chain, as shown in the right of Fig.\ref{fig:stitch}. The local rotation yields a dispersive coordinate $p_n$ and a compact Fano coordinate $f_n$. The sequence $\epsilon_n^+$ describes onsite perturbations of both $p_n$ and $f_n$, while the sequence $\epsilon_n^-$ couples the dispersive to the Fano coordinate within the rotated unit cell \cite{flach14}.  Solving for the Fano coordinates $f_n$ in the second equation above, we obtain a new equation for the dispersive portion 
\begin{equation}
(E+t)\, p_n= \left[\epsilon_n^+ + \frac{(\epsilon_n^-)^2}{(E-t) - \epsilon_n^+ 
}\right]\, p_n  - 2\left( p_{n-1} + p_{n+1} \right).
\label{eq:CS-pst}
\end{equation}
The reduced topology assumes the tight-binding form.
If eigenmodes are exponentially localized, their asymptotic decay is $\psi_n^\nu \sim e^{-\frac{n}{\xi}}$. The rate $\xi^{-1}(E)$ is the inverse localization length of a localized state at eigenenergy $E\in\mathbb{R}$, found by applying the recursive iteration
\begin{equation}
\xi^{-1}(E,\lambda) = \lim_{M\rightarrow+\infty}\frac{1}{M}\sum_{n=1}^M \ln\Big|  \frac{p_{n+1}}{p_n}\Big|.
\label{eq:loclength}
\end{equation}
for any given potential strength $\lambda$. We will use this method in all the 
numerical computations of the two models' localization lengths, for $M=10^6$. 
The energy $E$ in Eq.(\ref{eq:CS-pst})  will be numerically found from an 
exact diagonalization of a finite lattice of $N=512$ unit cells. 
In all the figures of the paper, if the recursive iteration converges to a finite value (chosen \cite{arbitrary} here as $\xi \le N/10$), the datapoint $(E,\lambda)$ is declared a localized state and plotted in blue. Otherwise if the iteration diverges, the datapoint is declared an extended state and plotted in red. 
\subsection{Symmetric Case: Metal-Insulator Transition}
We analyze first the symmetric case $\epsilon_n^-=0$, obtained for $\theta_b=0.0 $. Eq.(\ref{eq:CS}) reads
\begin{equation}
\begin{aligned}
(E+t)\, p_n &= \epsilon_n^+\, p_n - 2\left( p_{n-1} + p_{n+1} \right), \\
(E-t)\, f_n &= \epsilon_n^+\, f_n
\end{aligned}
\label{eq:CS-sym}
\end{equation}
with $\epsilon_n^+=\epsilon_n^i$.
The two sets of states $p_n$ and $f_n$ decouple and generate two independent spectra, respectively labeled $\sigma_p$ and $\sigma_f$. The parameter $t$ then simply operates as a shift parameter, translating $\sigma_p$ and $\sigma_f$ relative to each other by $2t$.

\begin{figure}[!ht]
 \centering
 \includegraphics[width=0.95\columnwidth]{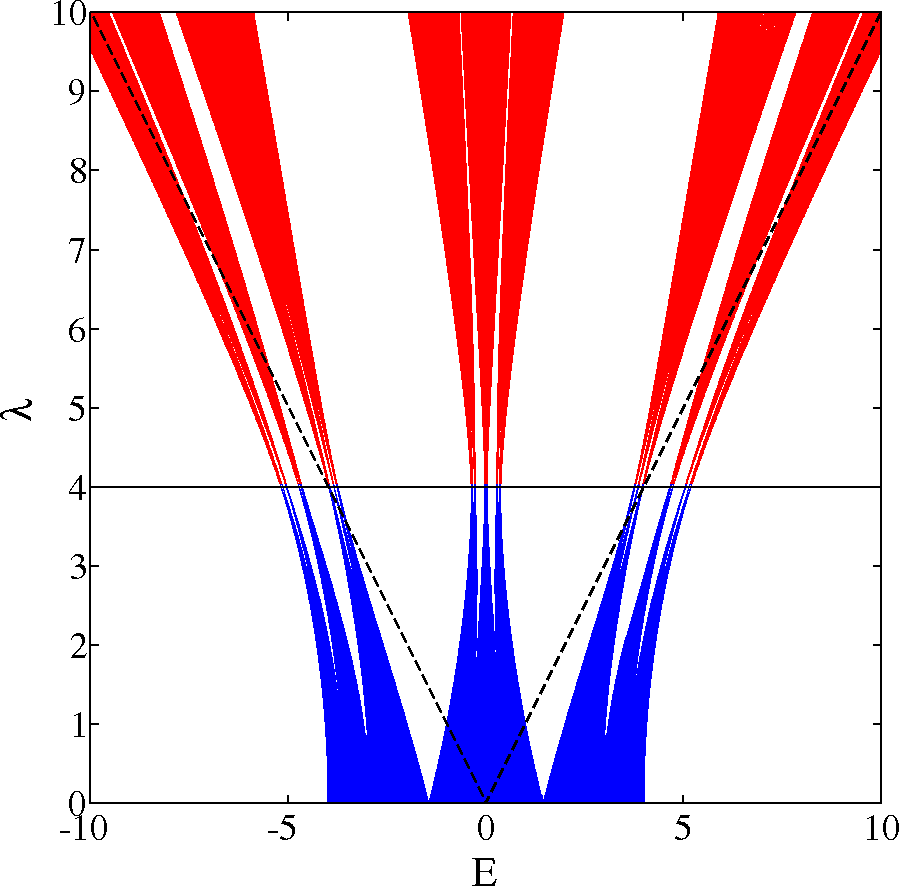}
 \caption{Symmetric Case:  The dispersive spectrum ($\sigma_p$) of the 
cross-stitch lattice, for $\epsilon_n^-=0$ and $t=0$. The Fano state spectrum 
$\sigma_f$ is omitted, but its boundaries indicated by black dashed lines. The 
black line represents the MIT at $\lambda=4$, which clearly separates extended 
states (blue) from those localized (red).}
 \label{Fig2}
\end{figure}

The dispersive states $p_n$ are described by an Aubry-Andr\'{e} chain, 
displaying a MIT at $\lambda_c=4$. 
The $\sigma_f$ states keep their compact feature, but the degeneracy of their eigenenergies is now removed -
these eigenenergies are given by $E=\epsilon_n^+ + t$. 
In Fig.\ref{Fig2}, we plot the spectrum from Eq.(\ref{eq:CS-sym}), as a 
function of $\lambda$. 
In this symmetric case, the spectra 
$\sigma_p$
and $\sigma_f$ 
are independent, since the $p_n$ and the $f_n$ coordinates decouple.

For every potential strength $\lambda>0$, the Fano states spectrum 
$\sigma_f=\{\epsilon_n^+ \}_n$ is equidistributed within the interval 
$[t-\lambda,t+\lambda]$. In Fig.\ref{Fig2} we indicate its boundaries by dashed 
lines. For the dispersive spectrum $\sigma_p$, the localization length is 
numerically found with the recursive iteration (\ref{eq:loclength}), and the 
localized phase (red) is demarcated from the extended one (blue). At the 
critical valuce $\lambda_c=4$, all dispersive states switch from extended to 
localized. In this case, there is no mobility edge.
%%%%%%%%%%%%%%%%%%%%%%%%%%%%%%%%%%%%%%%%%%%%%%%%%%%%%%%%%%%%%%%%%
%%%%%%%%%%%%%%%%%%%%%%%%%%%%%%%%%%%%%%%%%%%%%%%%%%%%%%%%%%%%%%%%%%
%%
%%
\subsection{Asymmetric Case $\epsilon_n^-\neq 0$: Numerical Evidence for Mobility Edge}  
\begin{figure}[!ht]
 \centering
 \includegraphics[width=0.95\columnwidth]{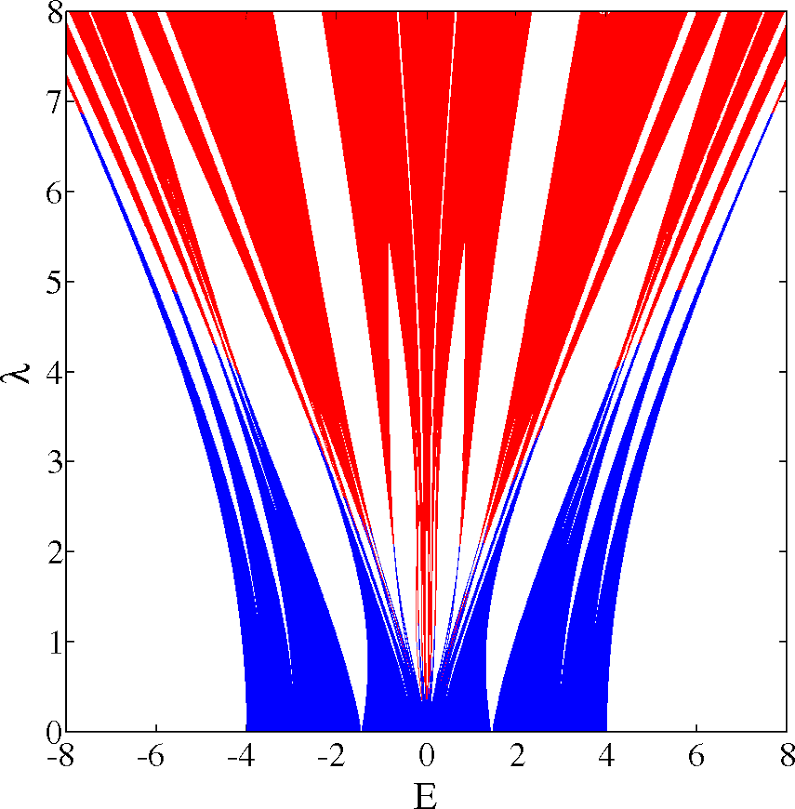}
 \caption{Asymmetric Case: Spectrum of the cross-stitch lattice, for 
$\theta=0.25$ and $t=0$. The extended (localized) portion of the spectrum is 
shown in blue (red), while the color boundary is a mobility edge 
approximation of the spectrum.}
 \label{Fig3}
\end{figure}

Breaking the symmetry $\epsilon_n^-\neq 0$ ($\theta_b \neq 0$) of the Fano chain Eq.(\ref{eq:CS}) effectively couples the dispersive states $p_n$ to their compact Fano counterparts $f_n$. Therefore,
the self-duality is lost, and the two independent spectra ($\sigma_{p,f}$) are 
now joint. Nevertheless, we expect a transition between localized and 
extended states via an energy-dependent mobility edge. In Fig.\ref{Fig3} we plot 
the spectrum of the lattice in the asymmetric case for $\theta_b=0.25$ and 
$t=0$. A mobility edge is clearly observed separating the localized regime (red) 
from that which is extended (blue).
%%%%%%%%%%%%%%%%%%%%%%%%%%%%%%%%%%%%%%%%%%%%%%%%%%%%%%%%%%%%%%%%%%%
%%%%%%%%%%%%%%%%%%%%%%%%%%%%%%%%%%%%%%%%%%%%%%%%%%%%%%%%%%%%%%%%%%%
%%
%%
\subsection{Antisymmetric Case ($\epsilon_n^+ = 0$): Analytic Mobility Edge}
\begin{figure}[!ht]
 \centering
 \includegraphics[width=0.95\columnwidth]{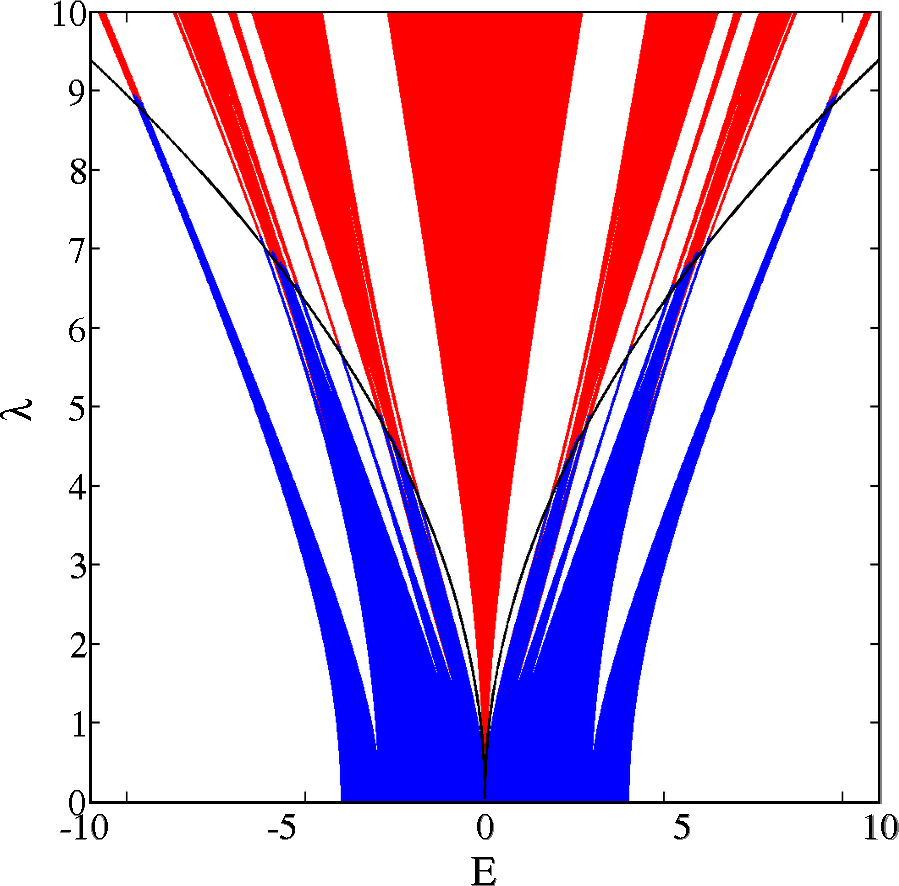}
 \caption{Antisymmetric Case: Spectrum of the cross-stitch lattice, for 
$\epsilon_n^+=0$ and $t=0$. The extended (localized) spectral portion is 
shown in blue (red). The color boundary is an approximation to the spectral 
mobility edge, however the black line is the analytical form of 
Eq.(\ref{eq:CS-ME}) -- a good agreement is observed.}
\label{Fig4}
\end{figure}

Among all the possible non-symmetric cases obtained for $\theta_b\neq 0,$ the antisymmetric one $\theta_b=0.5$ deserves special attention. In this situation $\epsilon_n^+=0$ and Eq.(\ref{eq:CS}) transforms into
\begin{equation}
\begin{aligned}
(E + t)\, p_n &= \epsilon_n^- f_n  - 2 (p_{n-1} + p_{n+1}), \\
(E - t)\, f_n &= \epsilon_n^-  p_n.
\end{aligned}
\label{eq:CS_pn_AS}
\end{equation}
It follows that all flatband states are expelled from their unperturbed energy position $E=t$. Since $\epsilon_n^-\neq 0$, from the second equation of (\ref{eq:CS_pn_AS}) it follows that at the flatband energy
\begin{equation}
\epsilon_n^- \,  p_n=0 \quad \Rightarrow \quad  p_n=0\ ,
\end{equation}
Then, from the first equation of (\ref{eq:CS_pn_AS}) we conclude
\begin{equation}
\epsilon_n^- \, f_n=0 \quad \Rightarrow \quad f_n=0.
\end{equation}
Only the trivial state $(p_n,f_n) = (0,0)$ satisifies Eq.(\ref{eq:CS_pn_AS}) exactly at the flatband energy $E=t$. In Fig.\ref{Fig4} we plot the spectrum for this antisymmetric case for $t=0$. We again observe a mobility edge. However, we can now even 
identify its analytical form, as observed in \cite{bodyfelt14}. Indeed the dispersive part, Eq.(\ref{eq:CS-pst}), reads
\begin{equation}
(E+t)\, p_n=\dfrac{(\epsilon_n^-)^2}{E - t} \, p_n- 2(p_{n-1} + p_{n+1})\;.
\label{disppor1}
\end{equation}
By trigonometric bisection
\begin{equation}
(\epsilon_n^-)^2 = \lambda^2\cos^2(2\pi\alpha n) = \dfrac{\lambda^2}{2}\left[1 + \cos(4\pi\alpha n)\right].
\label{eq:bisection}
\end{equation}
%%%%%%%%%%%%%%%%%%%%%%%%%%%%%%%%%%%%%%%%%%%%%%%%%%%%%%%%%%%%%%%
%%%%%%%%%%%%%%%%%%%%%%%%%%%%%%%%%%%%%%%%%%%%%%%%%%%%%%%%%%%%%%%
%%

Substituting Eq.(\ref{eq:bisection}) back into the previous equation, we obtain
\begin{multline}
\tilde{E} \; p_n = \dfrac{\lambda^2}{4(E-t)}\cos(4\pi\alpha n) - (p_{n-1} + 
p_{n+1}), \\
\mbox{where } \tilde{E} := \dfrac{E + t}{2} - \dfrac{\lambda^2}{4(E-t)}. \hfill
\end{multline}
Therefore, the model becomes an Aubry-Andr\'{e}  chain eigen-equation, but 
with onsite perturbation strength depending on $\lambda$ and $E$. From 
\cite{Aubry80}, the MIT occurs when the potential strength is twice larger than 
the hopping strength. Imposing that condition, an analytic expression is found 
for the mobility edge, $\lambda_c(E_c)$:
%%%
\begin{equation}
\abs{\dfrac{\lambda_c^2}{4(E_c-t)}} = 2 \quad \Rightarrow \quad \lambda_c(E_c)= 
2\sqrt{2|E_c-t|}\;.
\label{eq:CS-ME}
\end{equation}
Note that for $E_c=t$, the critical potential strength $\lambda_c$ vanishes in a 
square root manner, where the previously discussed lack of states at the 
flatband energy $E=t$ occurs. The analytic curve of the mobility edge is plotted 
in Fig.\ref{Fig4}, displaying excellent agreement with the numerical result.
\section{Diamond Lattice}\label{sec:diamond}
Under Eqs.(\ref{eq:fano-DC}, \ref{eq:eps+-}), the diamond lattice's 
Eq.(\ref{eq:fb-gen}) become
\begin{equation}
\begin{aligned}
(E+t)\, p_n\ &= \epsilon_n^+ p_n + \epsilon_n^- f_n - \sqrt{2} (c_n+c_{n+1}), \\
(E-t)\, f_n\ &= \epsilon_n^+\, f_n + \epsilon_n^-\, p_n, \\
(E - \epsilon_n^c)\, c_n &=  - \sqrt{2} (p_{n-1} + p_n),
\end{aligned}
\label{eq:DC}
\end{equation}
as illustrated graphically in the right plot of Fig.\ref{fig:diamond}. 

\begin{figure}[h!]
 \centering
 \includegraphics[width=0.95\columnwidth]{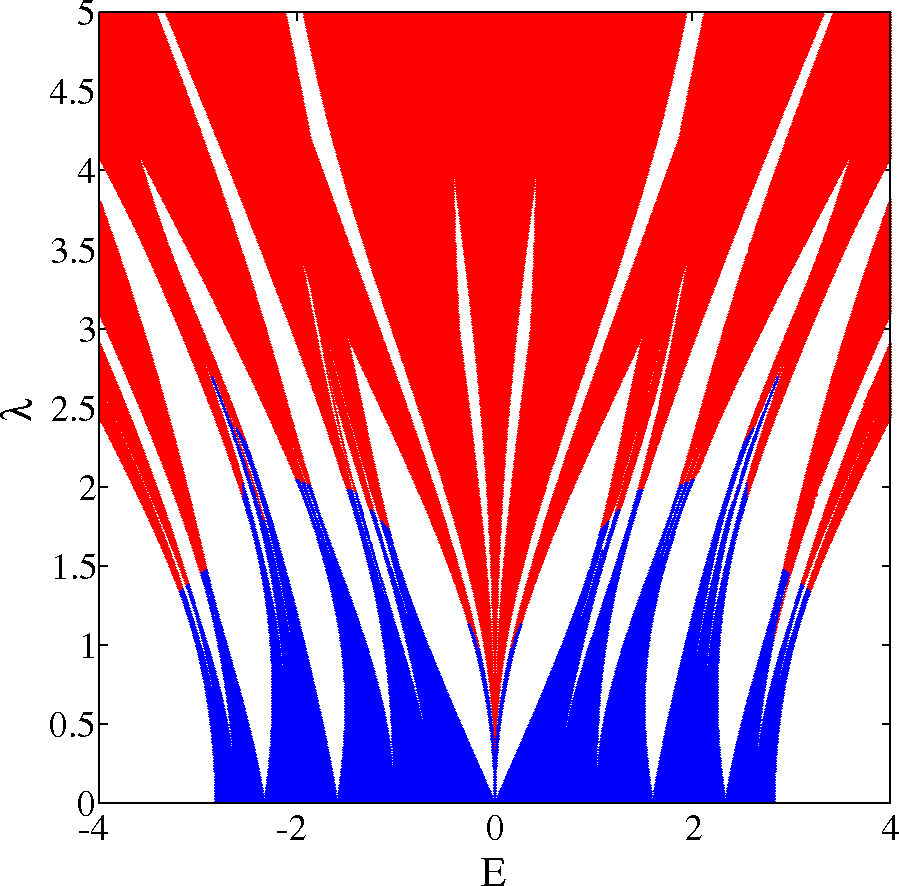}
 \caption{Spectrum of the diamond lattice, for  $\theta_b=0.5$, $\theta_c=0.05$ 
and $t=0$. The extended (localized) spectral portion is shown in blue (red), 
while the color boundary is again a spectral mobility edge approximation.}
 \label{Fig5}
\end{figure}

Expressing the $f_n$ and $c_n$ variables through the $p_n$ ones
we reduce these equations to a tight-binding form
which contains the $p_n$ coordinates only:
\begin{multline}
(E+t)\, p_n = \biggl[ \epsilon_n^+ + \frac{(\epsilon_n^-)^2}{(E-t) 
\epsilon_n^+ }  + \frac{2}{E-\epsilon_n^c} +  \frac{2}{E-\epsilon_{n+1}^c}  
\biggr] p_n  + \\
\frac{2}{E-\epsilon_n^c}p_{n-1} +\frac{2}{E-\epsilon_{n+1}^c} p_{n+1}. 
\label{eq:DC pn}
\end{multline}
We use this tight-binding form to numerically obtain the eigenvalue spectrum and consequently the localization length using the iterative recursion Eq.(\ref{eq:loclength}). In Fig.\ref{Fig5}, we plot the  diamond lattice spectrum and the mobility edge for $\theta_b=0.5$, $\theta_c=0.05$ and $t=0$.  

%%%%%%%%%%%%%%%%%%%%%%%%%%%%%%%%%%%%%%%%%%%%%%%%%%%%%%%%%%%%%%%%%%%%%%%%%%%%%%%%%%%%%%
%%%%%%%%%%%%%%%%%%%%%%%%%%%%%%%%%%%%%%%%%%%%%%%%%%%%%%%%%%%%%%%%%%%%%%%%%%%%%%%%%%%%%%

\begin{figure}[h]
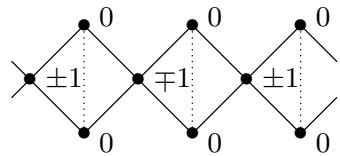

\scalebox{1.2}{
%\hspace{0mm}
\xy 
(32,0)*{\bullet} ;  (38,6)*{\bullet} ;  (38,-6)*{\bullet} ;(44,0)*{\bullet} ;  (50,6)*{\bullet} ;  (50,-6)*{\bullet} ; (56,0)*{\bullet} ;(62,6)*{\bullet} ;  (62,-6)*{\bullet} ;  
(38,6)*{} ; (50,-6)*{} **\dir{-} ; (50,6)*{} ; (62,-6)*{} **\dir{-} ;
 (38,-6)*{} ; (50,6)*{} **\dir{-} ; (50,-6)*{} ; (62,6)*{} **\dir{-} ; (38,-6)*{} ; (30,2)*{} **\dir{-} ; (38,6)*{} ; (30,-2)*{} **\dir{-} ; (62,6)*{} ; (66,2)*{} **\dir{-} ; 
(62,-6)*{} ; (66,-2)*{} **\dir{-} ;  
%(33,6)*{a} ; (33,-6)*{b} ; (27,0)*{c} ; (48,-11)*{n} ;  (36,-11)*{n-1} ; (60,-11)*{n+1} ; 
(38,6)*{} ; (38,-6)*{} **\dir{.} ; (50,6)*{} ; (50,-6)*{} **\dir{.} ; (62,6)*{} ; (62,-6)*{} **\dir{.} ;
(40.5,7)*{0} ; (40.5,-7)*{0} ; (52.5,7)*{0} ; (52.5,-7)*{0} ; (64.5,7)*{0} ; (64.5,-7)*{0} ; (35.8,0)*{\pm1} ; (47.8,0)*{\mp1} ; (59.8,0)*{\pm1} ; 
\endxy
}
\caption{Extended state $\mathcal{D}$ at energy $E=K$ on the diamond lattice in case of constant onsite potential $\epsilon_n^c = K$ on the $c$-chain (up to normalization).}
\label{stateD}
\end{figure}

We notice that in Eq.(\ref{eq:DC pn}) the coefficients of the hopping terms 
depend on the $c$-chain onsite perturbation $\epsilon_n^c$. Therefore, in 
general all hopping terms are perturbation-dependent. In order to arrive at an 
Aubry-Andr\'e form of  Eq.(\ref{eq:DC pn}) we set constant onsite energies on 
all $c$ sites: $\epsilon_n^c = K \in\mathbb{R}$. Then an extended state 
$\mathcal{D}$ exists at energy $E=K$, regardless of the other control 
parameters in Eq.(\ref{eq:DC}):
\begin{equation}
E=K\;,\; c_n=(-1)^n\;,\;f_n=p_n=0\;.
\label{Kstate}
\end{equation}
The state's amplitudes reside only on the $c$ sites (see Fig.\ref{stateD}).
The existence of this extended state is not affected by the perturbation strength $\lambda$, the flatband energy $E=t$, and any specific correlation of the onsite potential.
Therefore if the model admits a mobility edge curve $\lambda_c(E_c)$, it 
follows that it will diverge $\lambda_c(E_c=K) =\infty$, yielding a singularity.
\subsection{Symmetric Case: Analytic Mobility Edge}
We consider first the symmetric case $\epsilon_n^-=0$, obtained for $\theta_b=0.0$. In this situation, Eq.(\ref{eq:DC}) reads
\begin{eqnarray}
\nonumber
(E+t)\, p_n\ &=& \epsilon_n^+\, p_n +- \sqrt{2} (c_n+c_{n+1}) \;,\\
(E-t)\, f_n\ &=& \epsilon_n^+\, f_n \;, \\
\nonumber
(E - K)\, c_n &=&  - \sqrt{2} (p_{n-1} + p_n)\;.
\end{eqnarray}
The $f_n$ states decouple from both $p_n$ and $c_n$ states, generating two 
independent spectra $\sigma_f$ and $\sigma_{p,c}$. The flatband energy $t$ 
shifts the two energy spectra of $2t$ relative to each other. Substituting the 
$c$ variables by the $p$ ones we obtain the Aubry-Andr\'e form
%%%%%%%%%%%%%%%%%%%%%%%%%%%%%%%%%%%%%%%%%%%%%%%%%%%%%%%%%%%%%%%%%%%%%%%%%%%%%%%%
\begin{figure}[ht!]
\centering
\includegraphics[width=0.95\columnwidth]{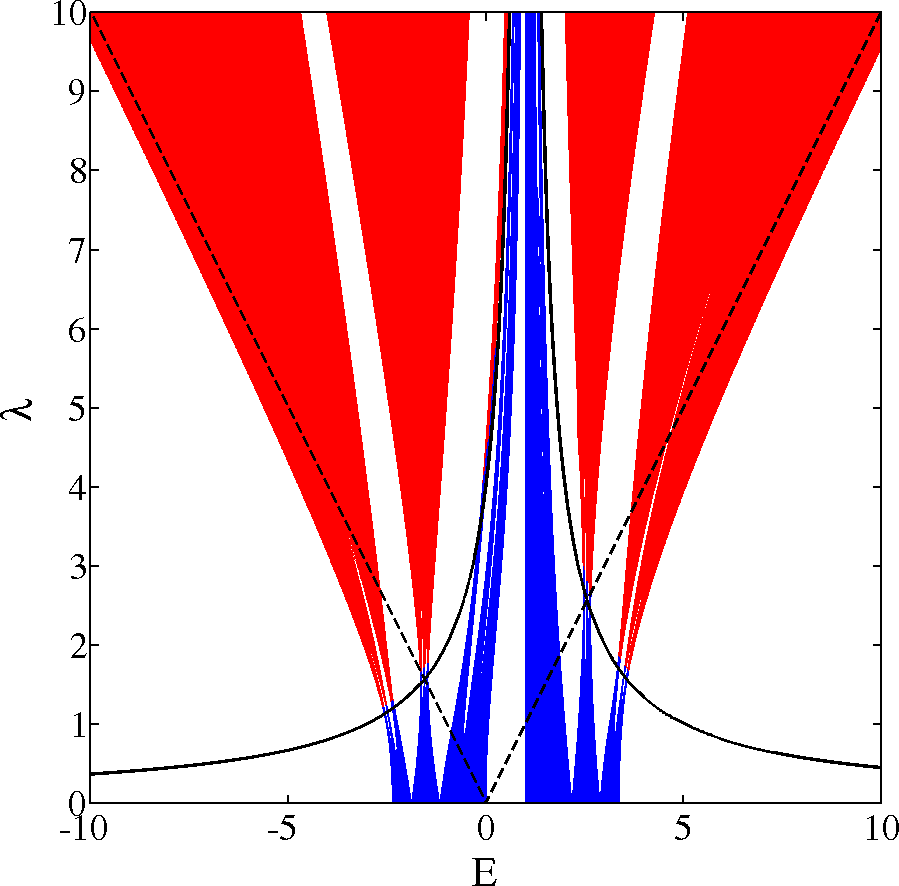}
\caption{Symmetric Case: Dispersive spectrum ($\sigma_{p,c}$) of the diamond 
lattice, for $\epsilon_n^+=0$, $K=1$, $t=0$. The Fano state spectrum 
($\sigma_f$) is omitted, but its boundaries are indicated by black dashed 
lines. The extended (localized) states of the dispersive spectrum are shown in 
blue (red) -- the boundary between these is a mobility edge approximation, 
in good agreement with its analytical form of Eq.(\ref{eq:ME_D}), shown as a 
solid black line. Note that at $E=K$, the mobility edge diverges to infinity.}
\label{Fig6}
\end{figure}
%%%%%%%%%%%%%%%%%%%%%%%%%%%%%%%%%%%%%%%%%%%%%%%%%%%%%%%%%%%%%%%%%%%%%%%%%%%%%%%%
%%%%%%
\begin{multline}
 \tilde{E}\, p_n = \frac{(E- K)\lambda}{2} \cos(2\pi\alpha n)\, p_n + p_{n-1} 
+p_{n+1},  \\
 \mbox{where } \tilde{E} :=\frac{ (E+t)(E- K)}{2} - 2. \hfill
\end{multline}
Note the onsite potential strength is now dependent on $\lambda$ and $E$. 
Imposing the equality between the potential strength and the Aubry-Andr\'e 
critical value, we arrive at the mobility edge
\begin{equation}
\abs{\frac{(E_c- K)}{2}\lambda_c} = 2 \quad \Rightarrow \quad \lambda_c(E_c) = \abs{\frac{4}{E_c-K}}\ .
\label{eq:ME_D}
\end{equation}
The mobility edge curve diverges at $E=K$ due to the existence of the delocalized state $\mathcal{D}$. We plot this mobility edge curve in Fig.\ref{Fig6} and observe very good agreement with the numerical results. 

\subsection{Antisymmetric Case: Analytic Mobility Edge} 
We consider the antisymmetric case $\epsilon_n^+= 0$ obtained with $\theta_b=0.5$, and with $\epsilon_n^c=K$. Eq.(\ref{eq:DC}) reads
\begin{equation}
\begin{aligned}
(E+t)\, p_n &=  \epsilon_n^- f_n - \sqrt{2} (c_n+c_{n+1})\;,\\
(E-t)\, f_n &=  \epsilon_n^- p_n \;, \\
(E - K)\, c_n &=  - \sqrt{2} (p_{n-1} + p_n)\;.
\end{aligned}
\label{eq:DC pn AS}
\end{equation} 
For $t\neq K$, all eigenenergies are expelled from the flatband energy $E=t$.
%
%
%%%%%%%%%%%%%%%%%%%%%%%%%%%%%%%%%%%%%%%%%%%%%%%%%%%%%%%%%%
\begin{figure}[!ht]
 \centering
 \includegraphics[width=0.95\columnwidth]{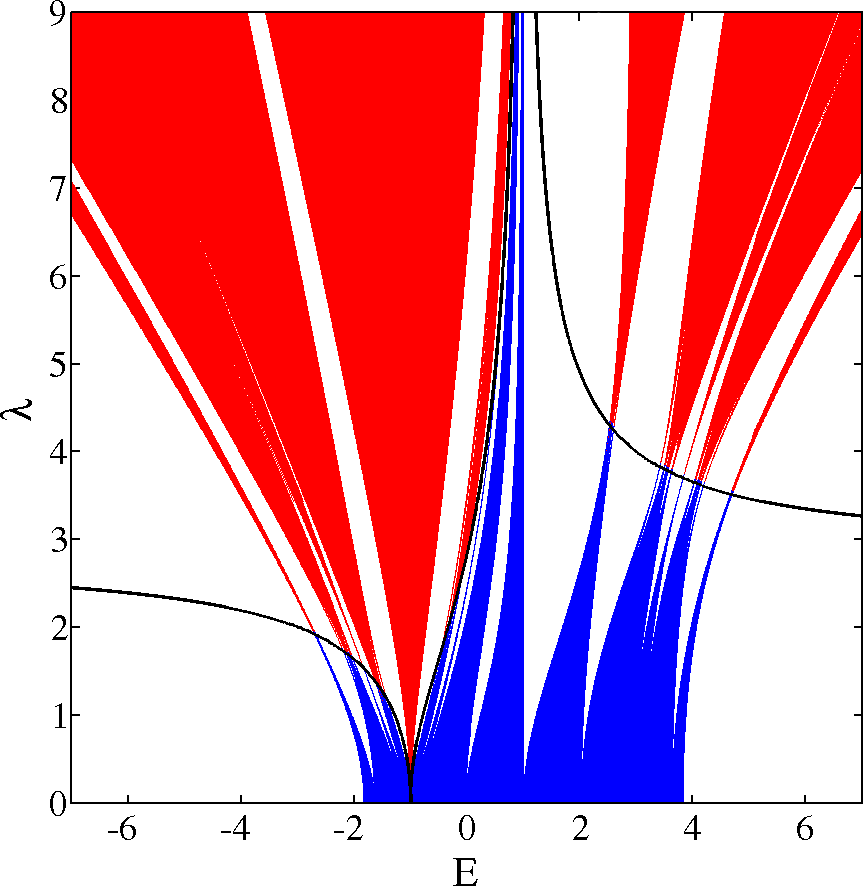}
 \caption{Antisymmetric Case: Spectrum of the diamond lattice, for 
$\epsilon_n^+=0$, $K=1$, $t=-1$. The extended (localized) spectral portion 
is shown in blue (red). The boundary between is an approximation of the 
spectrum's mobility edge -- in good agreement with the analytical form of 
Eq.(\ref{eq:ME_last}), plotted as a black line. Not the mobility edge 
curve zeroes at $E=t$, as well as diverges at $E=K$.}
\label{Fig7}
\end{figure}
%%%%%%%%%%%%%%%%%%%%%%%%%%%%%%%%%%%%%%%%%%%%%%%%%%%%%%%%%%%%%%%%%%%%%%%%%%%%%%%%
%
In Fig.\ref{Fig7} we plot the spectrum for this antisymmetric case for $t=-1$ 
and $K=1$. We derive an analytical expression of the mobility edge by reducing 
Eq.(\ref{eq:DC pn AS}) to an Aubry-Andr\'e form for the $p_n$ coordinates:
\begin{multline}
\tilde{E}\, p_n = \dfrac{E-K}{2} \dfrac{\lambda^2}{2(E - t)} \cos(4\pi\alpha 
n)\, p_n + p_{n-1} +p_{n+1}, \\
\mbox{where }\tilde{E}  := \dfrac{E- K}{2}\left[(E+t) - \dfrac{\lambda^2}{2(E 
- t)} \right] - 2. \hfill
\end{multline}
The condition for the MIT yields
\begin{equation}
\abs{\dfrac{E-K}{2} \dfrac{\lambda_c^2}{2(E - t)}} = 2 \; \Rightarrow 
\; \lambda_c(E) = 2\sqrt{2\abs{\dfrac{E-t}{E-K}}}\label{eq:ME_last}
\end{equation}
In Eq.(\ref{eq:ME_last}), the mobility edge curve $\lambda_c(E_c)$ diverges to 
infinity at $E=K$, in correspondence to the delocalized state $\mathcal{D}$. The 
curve also displays a zero at $E=t$, which corresponds to the lack of any states 
at the flatband energy \cite{LoS}. The mobility edge curve of 
Eq.(\ref{eq:ME_last}) is plotted in Fig.\ref{Fig7} and show excellent agreement 
with the numerical data. %%
\section{Conclusion}
Flatband topologies are characterized by macroscopic degeneracy at the flatband 
energy. General perturbations of these topologies lead to a removal of the 
degeneracy, yet keeps a high density of states and a bunching of the 
renormalized and hybridized states around the original flatband energy. This has 
especially dramatic consequencies for quasiperiodic Aubry-Andr\'e form 
perturbations. The flatband energy now hosts a zero of a mobility edge curve 
$\lambda_c(E_c)$. When approaching this zero the density of states grows, and 
the spatial extent of the eigenstates drops, making them more localized. For 
specific symmetries of the applied quasiperiodic potential, the dependence 
$\lambda_c(E_c)$ is obtained analytically, confirming the predicted zero, and 
further proving the strict nonexistence of any state at the former flatband 
energy. Some flatband topologies allow the existence of completely delocalized 
eigenstates at certain energies $E=K$. This leads to even more complex mobility 
edge curves which allow for a coexistence of zeroes and divergencies of 
$\lambda_c(E_c)$. Possible future topics of study include extension to $U>1$ 
topological classes and higher dimensional flatband models. It is our hope that 
the use of flatband topologies contributes interest to tunable mobility edges, 
e.g. by those realized in graphene~\cite{graphene}, monolayered 
dichalcogenides~\cite{chalcos}, or vandium dioxide films~\cite{VO2}. 

\newpage


\begin{thebibliography}{}
\bibitem{arbitrary}
We have varied this cutoff value up to $\xi \le N$, and it does not change the spectra drastically.
\bibitem{LoS}
With a similar argument discussed in Sec.\ref{sec:cross}, in this antisymmetry 
all the eigenstates are expelled from the flatband energy $E=t$.
\end{thebibliography}
\end{document}